# "Hot Spots" in Quasi-One-Dimensional Organic Conductors

Anatoley T. Zheleznyak and Victor M. Yakovenko

Department of Physics, University of Maryland, College Park, Maryland 20742-4111, USA

**Abstract**

The distribution of the electron scattering rate on the Fermi surface of a quasi-one-dimensional conductor is calculated for the electron-electron umklapp interaction. We find that in certain regions on the Fermi surface the scattering rate is anomalously high. The reason for the existence of these "hot spots" is analogous to the appearance of the van Hove singularities in the density of states. We employ a generalized $\tau$-approximation (where the scattering integral in the Boltzmann equation is replaced by the scattering time which depends on the position at the Fermi surface) to study the dependence of the electric resistance on the amplitude and the orientation of a magnetic field. We find that the "hot spots" do not produce a considerable magnetoresistance or commensurability effects at the "magic angles".

In a recent paper [1], Chaikin suggested phenomenologically that "hot spots", the regions where the electron scattering rate is anomalously high, may be present at the Fermi surface (FS) of the $(TMTSF)_2X$ organic metals. With a particular choice of the model parameters, he was able to explain the experimental data [2] on the angular dependence of magnetoresistance (the Lebed's "magic angles" effect [3]). However, the microscopic origin of the "hot spots" was not specified in Ref. [1]. In an earlier paper [4], it was mentioned that the electron-electron umklapp scattering or the scattering on a soft phonon mode in the vicinity of a charge/spin-density-wave transition may lead to formation of the "hot spots". The latter idea was elaborated in detail in Ref. [5]. In the present paper, we calculate straightforwardly the distribution of the electron scattering rate on the Fermi surface (FS) of a quasi-one-dimensional (Q1D) conductor due to the electron-electron umklapp scattering and demonstrate that the "hot spots" indeed exist.

Let us consider a Q1D metal which consist of parallel conducting chains. The axis $x$ is in the direction of the highest conductivity (along the chains), the axes $y$ (also referred to as $b$) and $z$ (or $c$) are in the directions of the lower and the lowest conductivities, and both directions are perpendicular to the chains. The Fermi momentum in the $x$-direction is equal to $\pm k_F$ ($\hbar = 1$). In the umklapp scattering process, the total momentum of the two scattering electrons changes by the value $4k_F$ which is equal to the lattice wave vector in the $(TMTSF)_2X$ compounds. The scattering rate (the inverse of the lifetime $\tau$) of the electron with the momentum $\vec{k} = (k_x, k_y, k_z)$ is given by the following general expression:

$$\frac{1}{\tau(\vec{k})} = g_4^2 \int \frac{d^3k_1\, d^3k_2\, d^3k_3}{(2\pi)^6} \delta[\vec{k} + \vec{k}_1 - \vec{k}_2 - \vec{k}_3]$$
$$n[\varepsilon_+(\vec{k})]\, n[\varepsilon_+(\vec{k}_1)]\, (1 - n[\varepsilon_-(\vec{k}_2)])\, (1 - n[\varepsilon_-(\vec{k}_3)])$$
$$\delta[\varepsilon_+(\vec{k}) + \varepsilon_+(\vec{k}_1) - \varepsilon_-(\vec{k}_2) - \varepsilon_-(\vec{k}_3)], \qquad (1)$$

where $g_4$ is the umklapp scattering amplitude, $n(\varepsilon)$ is the Fermi distribution function at the temperature $T$, and every integration is over the Brillouin zone (BZ).

In a Q1D conductor, the electron dispersion law can be written in the linearized with respect to $k_x$ form:

$$\varepsilon_\alpha(\vec{k}) = \alpha v_F(k_x - \alpha k_F) + 2t_b \cos(k_y + \alpha\varphi_1)$$
$$+ 2t_c \cos(k_z + \alpha\varphi_2). \qquad (2)$$

Here the energy $\varepsilon$ is counted from the Fermi level, the index $\alpha = \pm$ labels the electrons whose momenta along the chains are close to $\pm k_F$, $v_F$ is the Fermi velocity, $t_b$ and $t_c$ are the electron hopping integrals between the chains ($t_b \gg t_c$), the transverse momenta $k_y$ and $k_z$ are dimensionless and vary from 0 to $2\pi$, and $\varphi_1$ and $\varphi_2$ are the phases determined by the crystal structure of the $(TMTSF)_2X$ compounds [6].

We shall consider only the scattering of the electrons located at the FS ($\varepsilon_+(\vec{k}) = 0$). In this case, the scattering rate becomes a function of only two variables, $k_y$ and $k_z$, which label positions on the FS; the third momentum $k_x$ is determined by the condition that $(k_x, k_y, k_z)$ belongs to the FS. Substituting (2) into (1) and taking several integrals, we find:

$$\frac{1}{\tau(k_y, k_z, T)} = T^2 \frac{C}{4} \left(\frac{g_4}{2\pi v_F}\right)^2 G(k_y, k_z), \qquad (3)$$

$$G(k_y, k_z) = \int_0^{2\pi} \frac{dk_y^{(1)} dk_y^{(2)} dk_z^{(1)} dk_z^{(2)}}{(2\pi)^4}$$
$$f[t_b F(k_y, \varphi_1; k_y^{(1)}, k_y^{(2)}) + t_c F(k_z, \varphi_2; k_z^{(1)}, k_z^{(2)})], \qquad (4)$$

$$f(\zeta) = \frac{\zeta}{CT^2 \sinh(\zeta/T)}, \quad C = \int_{-\infty}^{\infty} \frac{\zeta\, d\zeta}{\sinh(\zeta)} = 4.9, \qquad (5)$$

$$F(p, \varphi; p_1, p_2) = \cos(p + \varphi) + \cos(p_1) + \cos(p_2)$$
$$+ \cos(p + p_1 - p_2 - 3\varphi). \qquad (6)$$

The normalizing coefficient $C$ is introduced in Eqs. (3) and (5) in such a way that the function $f(\zeta)$ becomes the Dirac function $\delta(\zeta)$ when $T \to 0$. One may conclude that at low temperatures the integration over momenta in (4) gives a temperature-independent geometrical factor $G(k_y, k_z)$, and, for all values of $(k_y, k_z)$, $1/\tau$ (3) is proportional to $T^2$. In other words, the scattering rate (3) is a factorized function of the temperature and the position at the FS in agreement with a standard Fermi liquid theory. However, as we shall see below, the factorization breaks in the two-dimensional (2D) case when $t_c = 0$ in Eq. (4).

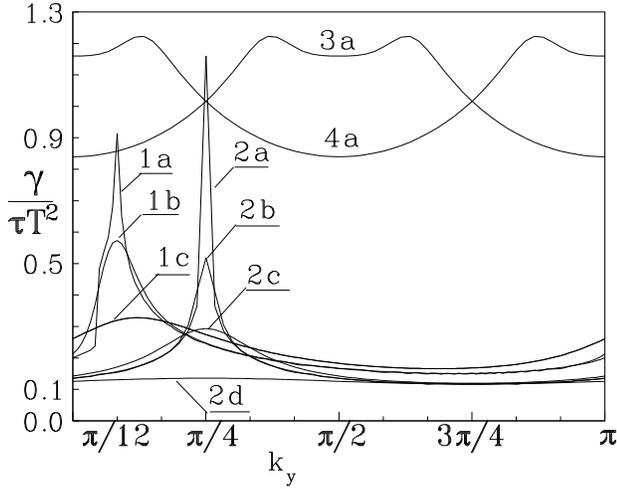
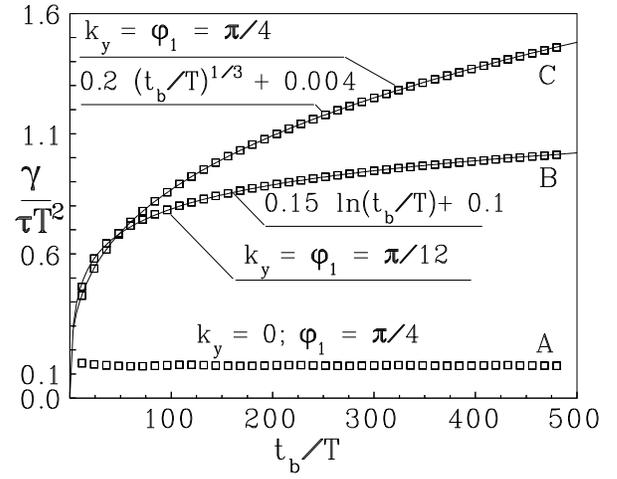

Figure 1: Variation of the electron scattering rate $1/\tau$ along the FS, 2D case. Numbers 1, 2, 3, and 4 correspond to the phases $\varphi_1 = \pi/12$, $\pi/4$, 0, and $\pi/2$; letters $a$, $b$, $c$, and $d$ define the temperature $T = t_b/240$, $t_b/24$, $t_b/5$, and $t_b$. The normalizing coefficient $\gamma$ is equal to $(2\pi)^4 t_b v_F^2 / 60 g_4^2$.

Figure 2: Temperature dependences of the electron scattering rates for different positions at the FS, 2D case. Squares represent the results of the numerical calculations, and the solid lines represent the corresponding curve fits. The coefficient $\gamma$ is the same as in Fig. 1.

Let us consider the 2D case first. In this case, $1/\tau$ is a function of only one variable $k_y$. Numerically calculated according to Eqs. (3)–(6) with $t_c = 0$ and normalized to $T^2$, the scattering rate $1/\tau(k_y)$ is shown in Fig. 1 for different temperatures $T$ and phases $\varphi_1$. Since Eq. (4) has the symmetry $G(k_y \pm \pi, k_z \pm \pi) = G(k_y, k_z)$, the results are shown only for a half of the BZ. We observe that, for certain special values of $k_y = k_y^*$, $1/\tau(k_y^*)T^2$ strongly increases as $T \to 0$. In other words, the points $k_y^*$ become increasingly "hotter" relative to other points on the FS as temperature decreases. Positions of these "hot points" are determined by the phase $\varphi_1$ in Eq. (2), namely $k_y^* = \varphi_1$. In the exceptional cases $\varphi_1 = 0$ and $\pi/2$, all points on the FS are "hot".

Temperature dependences of the scattering rate are shown in Fig. 2 for different positions at the FS. At the positions which are not "hot", the scattering rate is proportional to $T^2$ (curve A); whereas at the "hot points", $\tau^{-1} \sim T^2 \ln(t_b/T)$ when $\varphi_1 \neq \pi/4$ (curve B) and $\tau^{-1} \sim T^2 (t_b/T)^{1/3}$ when $\varphi_1 = \pi/4$ (curve C).

The origin of the "hot points" can be understood in the following way. In the 2D case, the geometrical factor $G(k_y)$ (4), can be written, in the limit $T = 0$, as:

$$G(k_y) = \int_0^{2\pi} \frac{dk_y^{(1)} dk_y^{(2)}}{(2\pi)^2} \delta[t_b F(k_y, \varphi_1; k_y^{(1)}, k_y^{(2)})]. \tag{7}$$

Integral (7) is divergent for certain values of the parameter $k_y$. This divergency is similar to the van Hove singularity in the density of states in the 2D case, because Eq. (7) is mathematically analogous to the expression for the density of states in the 2D case.

The characteristic function $F(k_y, \varphi_1; k_y^{(1)}, k_y^{(2)})$ (6), which enters Eq. (7), is shown in Fig. 3. As a function of the two variables $k_y^{(1)}$ and $k_y^{(2)}$, $F$ has six extrema, three of them being saddle points. At these points, the first derivatives of the function $F$ vanish: $\partial F / \partial k_y^{(1),(2)} = 0$. Positions of the saddle points depend on the values of the parameters $k_y$ and $\varphi_1$, and it is always possible to find such a value of $k_y = k_y^*$, that the saddle points (SP) appear at the zero level of the function $F$:

$$F(k_y^*, \varphi_1; SP) = 0. \tag{8}$$

In this case, integral (7) becomes logarithmically divergent in the same way as the density of states diverges at the van Hove singularity. It turns out that the values of the function $F$ at all three saddle points always coincide and are equal to: $F(k_y, \varphi_1; SP) = -2\sin(2\varphi_1)\sin(k_y - \varphi_1)$. Condition (8) is satisfied either when $\varphi_1 = 0$ or $\pi/2$, or when $k_y^* = \varphi_1$. These values are in agreement with the positions of the "hot points" found numerically (see Fig. 1).

When integral (7) is divergent, we should recall that originally it was not the Dirac $\delta$-function in this integral, but the function $f$ (5) which has a finite width of the order of $T$. The finite width cuts off the logarithmic divergency, so that $G(k_y^*) \sim \ln(t_b/T)$. In the special case $\varphi_1 = \pi/4$, all three saddle points merge together in the $(k_y^{(1)}, k_y^{(2)})$ space to form a "super"-saddle point where not only the first derivatives of the function $F$ vanish, but also the second derivatives do: $\partial^2 F / \partial k_y^{(i)} \partial k_y^{(j)} = 0$, $i,j = 1,2$. In this case, the expansion of the function $F$ in the powers of $k_y^{(i)}$ in the vicinity of the "super"-saddle point starts from the third power, thus the temperature cut-off in Eq. (7) produces a stronger dependence: $G(k_y^*) \sim (t_b/T)^{1/3}$. These results are in agreement with the numerical analysis (see Fig. 2).

Let us return now to the three-dimensional (3D) case when $t_c \neq 0$ in Eq. (4). The numerically calculated scattering rate in the 3D case is shown in Fig. 4 as a function of the position on the FS. One can notice the warped "hot lines". Their location on the FS is determined by the condition that the saddle points of the function $t_b F(k_y, \varphi_1; k_y^{(1)}, k_y^{(2)}) +$ $t_c F(k_z, \varphi_2; k_z^{(1)}, k_z^{(2)})$ (see Eq. (4)), defined in the four-dimensional (4D) space $(k_y^{(1)}, k_y^{(2)}, k_z^{(1)}, k_z^{(2)})$, are located at the zero level:

$$t_b \sin(2\varphi_1) \sin(k_y - \varphi_1) + t_c \sin(2\varphi_2) \sin(k_z - \varphi_2) = 0. \quad (9)$$

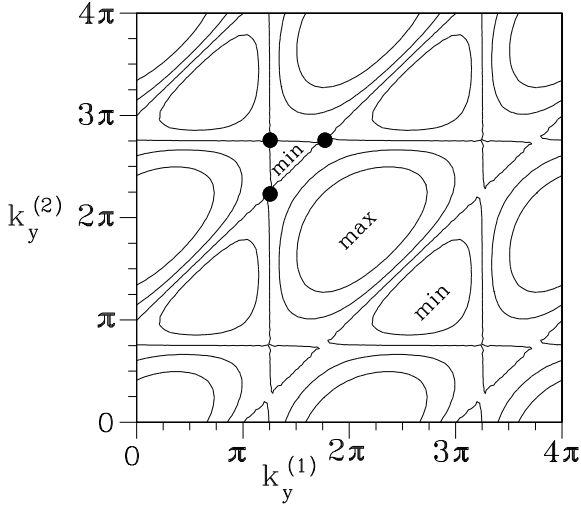

Figure 3: Characteristic function of the scattering rate, $F(k_y, \varphi_1; k_y^{(1)}, k_y^{(2)})$, shown as a function of $(k_y^{(1)}, k_y^{(2)})$ at $k_y = \varphi_1 = \pi/8$. Four BZ cells are shown for clarity. Positions of the saddle points are marked by black circles.

Saddle points in the 4D space, unlike in the 2D space, are integrable, that is they do not produce singularity in the function $G(k_y, k_z)$ (4). A non-zero value of $t_c$ ($t_c \ll t_b$) cuts off the 2D saddle point divergency, so the ratio between the crest and the valley in Fig. 4 is proportional to $\ln(t_b/t_c)$ (when $\varphi_1 \neq 0, \pi/4, \text{and} \pi/2$) or $(t_b/t_c)^{1/3}$ (when $\varphi_1 = \pi/4$). The variation of the scattering rate along the "hot lines" depends on the phase $\varphi_2$, but never exceeds 15%. At low temperatures $T \ll t_c$, the distribution of the scattering rate over the FS stays constant, and $\tau^{-1} \sim T^2$ at all points of the BZ.

Having calculated the scattering rate, let us calculate now the magnetoresistance. The linearized stationary kinetic equation in the magnetic field $H$ has the form [7]:

$$\frac{eHv_F}{c}\frac{\partial \psi}{\partial k_\perp} + \frac{\psi}{\tau(k_\parallel, k_\perp)} = e\vec{E}\cdot\vec{v}. \quad (10)$$

Here $n = n_0 - \psi \, \partial n_0/\partial \varepsilon$ is the electron distribution function, $n_0$ is the unperturbed Fermi function, the momenta $k_\parallel$ and $k_\perp$ are parallel and perpendicular to the magnetic field, $\vec{E}$ is the electric field, and $\vec{v} = \partial \varepsilon/\partial \vec{k}$ is the electron velocity. Dependence of the relaxation time $\tau$ on the variables $k_\parallel$ and $k_\perp$ in Eq. (10) reflects the variation of the scattering rate along the BZ. Solving Eq. (10) analytically and using the result to calculate the electric current, we find the expression for the electric resistivity along the chains, $\rho_{xx}$:

$$\frac{1}{\rho_{xx}} = \frac{2ec}{(2\pi)^3 H}\iint_{BZ} dk_\parallel \, dk_\perp$$
$$\int_0^\infty dk'_\perp \exp\left(-\frac{c}{eHv_F}\int_0^{k'_\perp}\frac{k''_\perp}{\tau(k_\parallel, k_\perp - k''_\perp)}\right). \quad (11)$$

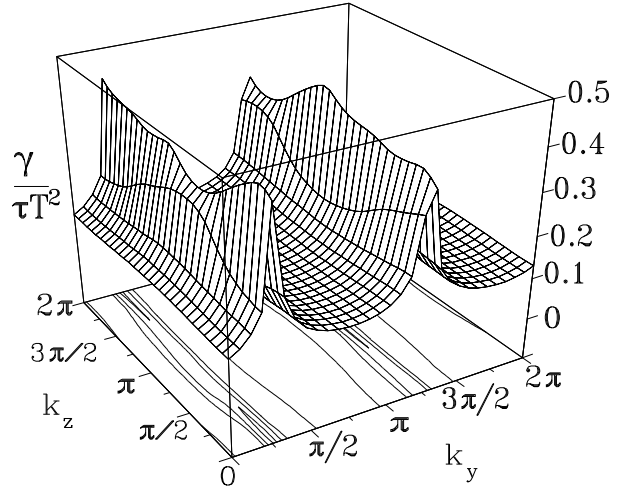

Figure 4: Variation of the electron scattering rate along the FS, 3D case, $\varphi_1 = \pi/4$, $\varphi_2 = \pi/12$, $t_b/t_c = 10$, and $t_b/T = 48$. The coefficient $\gamma$ is the same as in Fig. 1.

The orientation of the magnetic field implicitly enters Eq. (11) through the definitions of $k_\parallel$ and $k_\perp$ relative to the BZ.

The dependence of the resistance on the magnetic field $H \| c$ is shown in Fig. 5. It is natural to describe the value of the magnetic field using the parameter $h$ which shows how many times electrons cross the BZ before being scattered:

$$h = \frac{H}{H_0} = \frac{\omega_c \langle \tau^{-1}\rangle^{-1}}{2\pi}, \quad H_0 = \frac{2\pi c \langle \tau^{-1}\rangle}{bev_F}, \quad (12)$$

where $\langle ... \rangle$ means the averaging over the whole BZ, and $\omega_c = bev_F H/c$ is the cyclotron frequency. We observe from Fig. 5 that the resistance $\rho_{xx}(H)$ saturates at $h \geq 1$, when electrons cross the whole BZ at least once during their lifetime. The following relation between the limiting values of the resistance can be derived from Eq. (11):

$$\rho_{xx}(H=\infty)/\rho_{xx}(H=0) = \langle \tau \rangle / \langle \langle \tau^{-1}\rangle_{k_\perp}^{-1}\rangle_{k_\parallel}, \quad (13)$$

where $\langle ... \rangle_{k_{\parallel,\perp}}$ is the average taken in the direction parallel ($\|$) or perpendicular ($\perp$) to the magnetic field over the BZ. In the 2D case, when $H \| c$, Eq. (13) takes a simpler form:

$$\rho_{xx}(H=\infty)/\rho_{xx}(H=0) = \langle \tau \rangle \langle 1/\tau \rangle. \quad (14)$$

Due to the small variation of the scattering rate along the $k_z$ axis (see Fig. 4), the saturation value of the magnetoresistance in Fig. 5 is in agreement with the 2D formula (14). The calculated magnetoresistance is too small (13%) compared to the experiment [2] which shows a much higher value and no signs of saturation in a high magnetic field.

Since in the 3D case the scattering rate has the uniform temperature dependence $T^2$ in the low temperature region ($T \ll t_c$), the ratio $\rho_{xx}(H=\infty)/\rho_{xx}(H=0)$ (14) does not depend on temperature in this region. However, in the 2D case, the different temperature dependence of $\tau^{-1}(k_y^*)$ at the "hot points" can produce an increase of the magnetoresistance as $T \to 0$. In Fig. 6, we show the maximum value of magnetoresistance as a function of the inverse temperature for the different phases $\varphi_1$ in the 2D case. We observe that even in the "super"-saddle case ($\varphi_1 = \pi/4$), the temperature dependence of the magnetoresistance is weak and practically negligible.

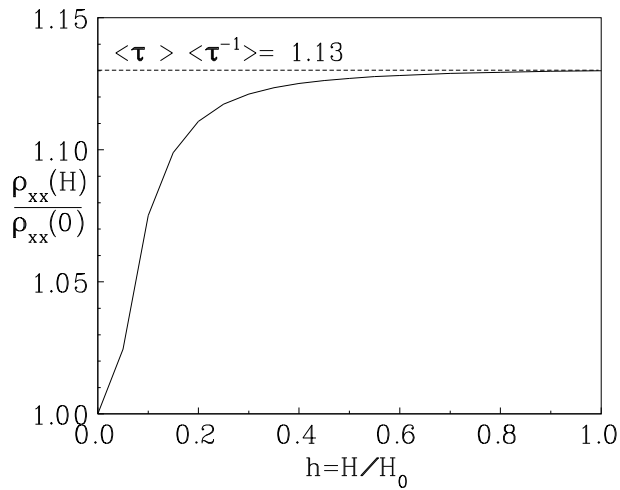

Figure 5: Magnetoresistance as a function of the magnetic field $H\|c$, 3D case, $H_0 = 2\pi c \langle \tau^{-1} \rangle / b e v_F$.

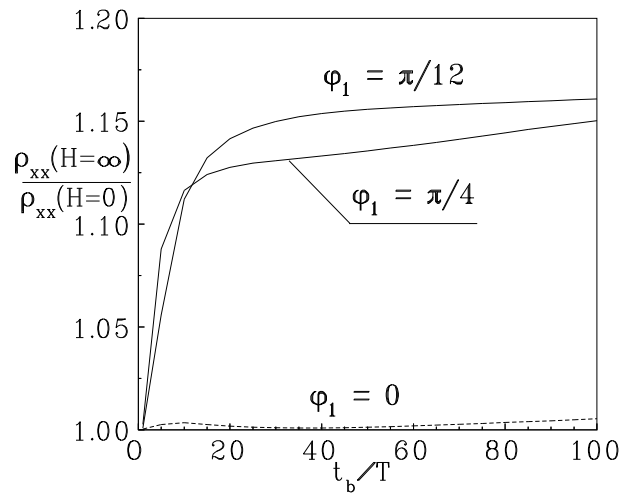

Figure 6: Temperature dependence of the saturation value of the magnetoresistance, 2D case.

In Fig. 7, the dependence of the magnetoresistance on the inclination angle of the magnetic field, $\theta$, is shown in the situation when the magnetic field is rotated perpendicular to the chains. In agreement with the experiment [2], Fig. 7 shows no magnetoresistance when $H\|c$, because in this orientation the Lorentz force does not sweep electrons into the "hot lines". However, in contradiction with the experiment, Fig. 7 shows no dips or any features at magic angles [3] when $\tan\theta$ is equal to a rational number. The curves in Fig. 7 turn out to be smooth because the scattering rate varies too weakly in the $k_z$ direction in Fig. 4.

We conclude that a strong variation of the electron scattering rate along the Fermi surface should exist in Q1D conductors in the case of electron-electron scattering. The reason for the variation is very general and analogous to the reason for the existence of the van Hove singularities in the density of states. However, we found that the variation is not strong enough to produce the high magnetoresistance or the angular commensurability effects observed experimentally. In our consideration we neglected the influence of the magnetic field on the scattering rate. That is correct in the classical region $T \gg \omega_c$. However, at low temperatures the influence, which is a quantum effect, must be taken into account as it was done in Ref. [3]. We conclude that the classical effect of the magnetic field, that is, a specific order of averaging of the scattering rate due to the Lorentz orbital motion of the electrons, is not enough to explain the peculiarities of the transport properties of the $(TMTSF)_2X$ conductors in a high magnetic field.

## References

[1] P. M. Chaikin, Phys. Rev. Lett. **69**, 2831 (1992).

[2] W. Kang, S. T. Hannahs, and P. M. Chaikin, Phys. Rev. Lett. **69**, 2827 (1992).

[3] A. G. Lebed and P. Bak, Phys. Rev. Lett. **63**, 1315 (1989); A. G. Lebed, J. Phys. (Paris) **4**, 351 (1994).

[4] C. S. Jacobsen, K. Mortensen, M. Weger, and K. Bechgaard, Sol. State Comm. **38**, 423 (1981).

[5] E. N. Dolgov and E. S. Nikomarov, Fiz. Tv. Tela **32**, 133 (1990) [Sov. Phys. Solid State **32**, 74 (1990)]; E. N. Dolgov, Fiz. Tv. Tela **31**, 23 (1989) [Sov. Phys. Solid State **31**, 1486 (1989)].

[6] K. Yamaji, J. Phys. Soc. Jpn. **55**, 860 (1986); A. G. Lebed, Physica Scripta. **39**, 386 (1991).

[7] A. A. Abrikosov, *Fundamentals of the Theory of Metals* (North-Holland, Amsterdam, 1988), Ch. 5.

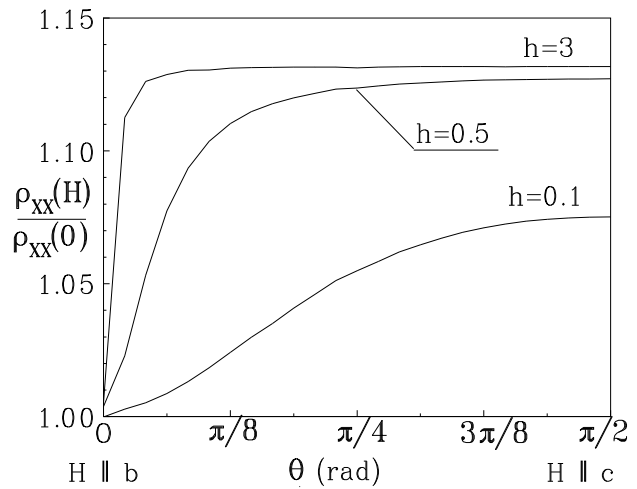

Figure 7: Angular dependence of the magnetoresistance for different values of the magnetic field, 3D case. The definition of $h$ is given in Eq. (12).